\documentclass[apj, iop]{emulateapj-rtx4}
\usepackage{natbib}

\shorttitle{Recombining Plasma in N49}
\shortauthors{Uchida et al.}

\begin{document}

\title{N49: the first robust discovery of a recombining plasma in an extra galactic supernova remnant}

\author{
Hiroyuki Uchida\altaffilmark{1},
Katsuji Koyama\altaffilmark{1,2}, 
Hiroya Yamaguchi\altaffilmark{3,4}
}

\altaffiltext{1}{Department of Physics, Graduate School of Science, Kyoto University, Kitashirakawa Oiwake-cho, Sakyo-ku, Kyoto 606-8502, Japan}
\email{uchida@cr.scphys.kyoto-u.ac.jp}
\altaffiltext{2}{Department of Earth and Space Science, Graduate School of Science, Osaka University, 1-1 Machikaneyama, Toyonaka, Osaka 560-0043, Japan}
\altaffiltext{3}{NASA Goddard Space Flight Center, Code 662, Greenbelt, MD 20771, USA}
\altaffiltext{4}{Department of Astronomy, University of Maryland, College Park, MD 20742, USA}

\begin{abstract}

Recent discoveries of recombining plasmas (RPs) in supernova remnants (SNRs) have dramatically changed our understanding of SNR evolution. To date, a dozen of RP SNRs have been identified in the Galaxy. Here we present Suzaku deep observations of four SNRs in the Large Magellanic Cloud (LMC), N49, N49B, N23, and DEM~L71, for accurate determination of their plasma state. Our uniform analysis reveals that only N49 is in the recombining state among them, which is the first robust discovery of a RP from an extra-galactic SNR. Given that RPs have been identified only in core-collapse SNRs, our result strongly suggests a massive star origin of this SNR. On the other hand, no clear evidence for a RP is confirmed in N23, from which detection of recombination lines and continua was previously claimed. Comparing the physical properties of the RP SNRs identified so far, we find that all of them are categorized into the ``mixed-morphology'' class and interacting with surrounding molecular clouds. This might be a key to solve formation mechanisms of the RPs.

\end{abstract}

\keywords{ISM: abundances --- ISM: individual (DEM~L71, N23, N49, N49B) --- supernova remnants --- X-rays: ISM}

\section{Introduction}

X-ray observations of supernova remnants (SNRs) allow us to make accurate measurements of 
plasma conditions and elemental abundances in the supernova (SN) ejecta, providing unique insights 
into the progenitor's evolution and explosion as well as the dynamical evolution of the remnants themselves. 
The hot plasmas in X-ray-emitting SNRs are commonly in a state of non-equilibrium ionization (NEI), 
where the ionization degrees of heavy elements are inconsistent with those expected for a collisional 
ionization equilibrium (CIE) plasma with a certain electron temperature \citep[e.g.,][]{Masai1984}.  
It had been widely believed for a long time that the SNR plasma is always in low-ionization, 
and slowly ionizing to achieve the CIE (hereafter ``ionizing plasma''; IP). 
In fact, recent X-ray observations with sensitive satellites, like {\it Suzaku} \citep{Mitsuda2007}, 
have confirmed that the immediate post-shock gas in young SNRs indeed consists of extremely 
low ionized atoms together with hot electrons \citep[e.g.,][]{Uchida2013,Yamaguchi2014Tycho}. 

However, earlier {\it ASCA} observations had suggested presence of a recombining plasma 
(hereafter RP, where the atoms are overionized compared to the observed electron temperature) 
in a couple of SNRs, IC~443 and W49B \citep{Kawasaki2002,Kawasaki2005}. 
The conclusive evidence for the RPs was revealed by later {\it Suzaku} observations 
that discovered enhanced radiative recombination continua (RRCs) in the X-ray spectra of 
these remnants \citep{Yamaguchi2009,Ozawa2009}. 
These results were followed by observational studies of other SNRs, leading to a significant 
increase in the number of RP SNRs. To date, RPs have been discovered from a dozen of SNRs, 
which means that the presence of RPs is no longer unusual or surprising.

The well-identified RP SNRs are all categorized into the so-called mixed-morphology (MM) class, 
defined by centrally-filled thermal X-ray emission with a synchrotron radio shell \citep{Rho1998}. 
While more than 25\% of the X-ray-detected Galactic SNRs are classified into this type \citep{Jones1998}, 
a physical process to form such characteristic morphology is still unclear.
It is theoretically pointed out that dense ambient materials play an important role in the dynamical evolutions 
of MM SNRs \citep[e.g.,][]{White1991,Petruk2001}. 
Notably, there is another prediction that the formation of RP can also be explained by interaction between the SN ejecta and dense 
materials surrounding the progenitor \citep[e.g.,][]{Itoh1989}. Therefore, physical processes that 
create the mixed-morphology and RP seem related with each other.

It is noteworthy that the most RP SNRs identified so far were previously considered to have an IP or a nearly-CIE plasma \citep[e.g., W44;][]{Uchida2012W44}. 
This implies that there are still a number of SNRs of which plasma state has been misclassified. 
Here, we present uniform analysis of high-quality data of MM SNRs (as well as typical shell-like SNRs simultaneously observed) in the Large Magellanic Cloud (LMC) obtained by the X-ray Imaging 
Spectrometer \citep[XIS;][]{Koyama2007} on board {\it Suzaku}, in order to search for a RP in these SNRs. 
The LMC is particularly suitable for such systematic study because of its low foreground extinction 
\citep{Dickey1990} and its known distance of 50\,kpc \citep{Feast1999}; we use this distance throughout the paper. 

In \S2, we outline our observations and data reduction procedures. The results of spectral analysis 
of four LMC SNRs, N49, N23, N49B, and DEM~L71 are presented in \S3 --- where we will show 
a new discovery of a RP from N49. We discuss our results in \S4, and finally give conclusions and 
future prospect in \S5. 
The errors quoted in text, tables, and figures are at the 90\% confidence level unless otherwise stated.

\begin{table*}[!t]
\caption{Observation logs.}\label{tab:obs}
\begin{center}
\begin{tabular}{lcccc}
\hline
\hline
Target & Obs. ID  & Obs. Date & (R.A., Decl.)$_{\rm{J2000}}$  & Exposure\\
\hline	   						    		      			  						
N23 \& DEM\,L71 & 807008010   & 2012 April 4 &  (76.45, --67.96)   & 102\,ks\\
N49 \& N49B  & 807007010  & 2012 May 9 &  (81.50, -66.08)   & 185\,ks\\
\hline
\end{tabular}
\end{center}
\end{table*}

\section{Observations and Data Reduction}

We observed two regions in the LMC aiming at the MM SNRs N23 and N49 during the {\it Suzaku} Cycle 8 phase. 
Detailed information of the observations is summarized in Table~\ref{tab:obs}.
In addition to the targeted sources, nearby SNRs DEM~L71 and N49B are detected in each field of view (FoV) of the XIS. 
We analyzed their spectra as well, for comparison of the plasma states among the targets. 
Both DEM~L71 and N49B are classified as standard shell-like SNRs \citep{Hughes2003, Park2003N49B} from which a recombining plasma has never been observed to date.

We analyzed data from three active CCDs; one is back-illuminated (XIS1) and the others are front-illuminated (XIS0 and XIS3), although only the merged and averaged front-illuminated (FI) CCD spectra are shown in Figures in the subsequent sections.
We used HEAsoft tools version 6.11 for the data reduction. 
The uncleaned data were reprocessed using the calibration database released in September 2011, and screened with the standard event selection criteria for cleaned events.
The resulting effective exposures were 102\,ks and 185\,ks for the observations of N23 (with DEM~L71) and N49 (with N49B), respectively.

\section{Analysis and Results}

\begin{figure}[t]
 \begin{center}
\includegraphics[width=85mm]{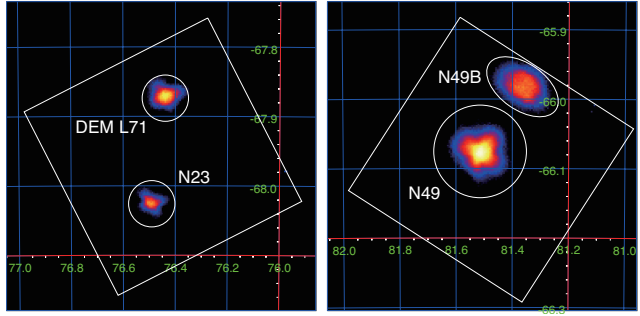}
 \end{center}
 \caption{
 Vignetting-corrected XIS FI images of the N23 (left) and N49 (right) regions in the 0.6--10.0~keV band with an equatorial coordinate grid. 
DEM L71 and N49B are also detected in the FoV of the XISs indicated by the solid squares.}
 \label{fig:image}
\end{figure}

We show in Figure~\ref{fig:image} the 0.6--10.0~keV images of the two observations after subtraction of non X-ray background (NXB) generated using \textit{xisnxbgen} \citep{Tawa08}.
Since the diameters of the SNRs \citep[$\lesssim 1\arcmin.5$; e.g., N49B;][]{Park2003N49B} are smaller than the angular resolution of the X-Ray Telescopes of  (XRTs; the half-power
diameter ranges from $1\arcmin.8$ to $2\arcmin.3$), our analysis focuses on spectra from the entire SNRs.
We extract the source spectra from the regions enclosed by the circles shown in Figure~\ref{fig:image}.
The background spectra are taken from the entire FoV excluding the source regions as well as the CCD corners illuminated by the $^{55}$Fe calibration sources.

\begin{figure}[t]
 \begin{center}
 \includegraphics[width=80mm]{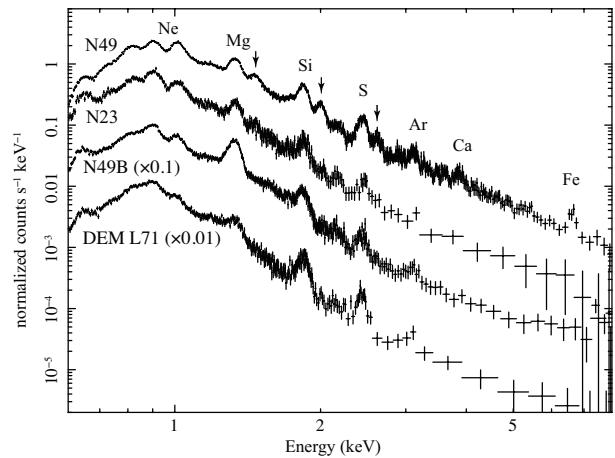}
 \end{center}
 \caption{XIS FI spectra of N49, N23, N49B and DEM~L71, where the background spectra are
subtracted. The spectra of N49B and DEM~L71 are multiplied by factors of 0.1 and 0.01,
respectively. The arrows represent the line centroids of Mg-Ly$\alpha$, Si-Ly$\alpha$
and S-Ly$\alpha$.}
 \label{fig:allspec}
\end{figure}

Figure~\ref{fig:allspec} compares background-subtracted spectra of the four SNRs, showing that N49 is the brightest among them (with the count rate about $\sim$10 times higher than
the others).
Owing to the good energy resolution of the XIS, we clearly detect K-shell emission lines of Ar, Ca, and Fe from this remnant, for the first time. The spectrum of N49 exhibits strong Ly$\alpha$ lines of Mg, Si and S, whereas there is little or no signal of these lines from the other remnants.
This indicates that the heavy elements in N49 are more highly ionized than those in the others.
As our main goal is to identify a RP SNR(s), we first perform detailed spectral analysis of N49 (\S3.1), which is followed by analysis of the other remnants (\S3.2--3.4).

We use the SPEX software version 2.04.01 \citep{Kaastra1996} for the spectral fitting by taking into account the detector and telescope responses (so-called ``redistribution matrix'' and
``ancillary response'', respectively) generated by \textit{xisrmfgen} and \textit{xissimarfgen} \citep{Ishisaki07}.
The data around the neutral Si K-shell edge (1.77--1.83\,keV) are ignored because of the poor accuracy in the response function at these energies\footnote{http://heasarc.nasa.gov/docs/Suzaku/analysis/sical.html}.
During the analysis, we separately set absorption column densities in the Milky Way ($N_{{\rm H(MW)}}$) and the LMC ($N_{{\rm H(LMC)}}$). The former value is fixed to $6\times10^{20}$~cm$^{-2}$ \citep{Dickey1990}.
Elemental abundances of the latter component are fixed to the averages values of the LMC \citep[$\sim$0.3\,solar;][]{Russell1992}.

\begin{figure}[t]
 \begin{center}
\includegraphics[width=85mm]{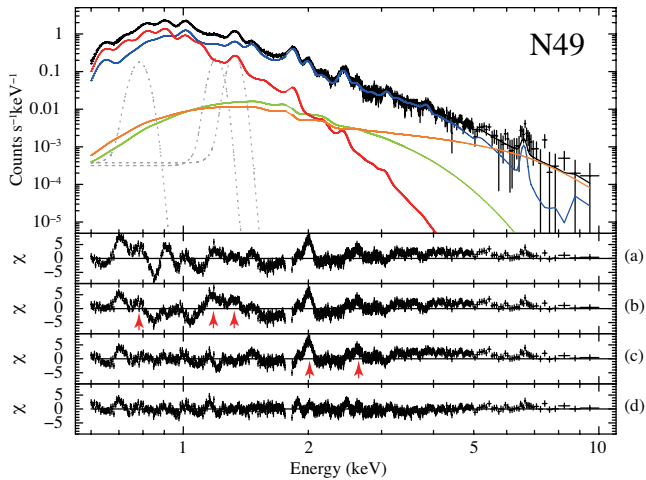}
 \end{center}
 \caption{XIS FI spectrum of N49 (top panel). The best-fit model is overlaid with
the black solid line. The solid blue and red lines represent the best-fit high-$kT_ e$
(RP) and the low-$kT_ e$ (IP) components, respectively. The orange and green lines
show the power-law and the blackbody components for SGR~0526$-$66, respectively.
The dotted lines show Gaussians for the Fe L emission missing from the plasma code we used.
The residuals are shown in
panel-(d). Panels (a), (b), and (c) represent the residuals from the models of single IP, two IP, and two IP plus Gaussians, respectively.}
 \label{fig:N49}
\end{figure}

\begin{figure*}[t]
 \begin{center}
 \includegraphics[width=70mm]{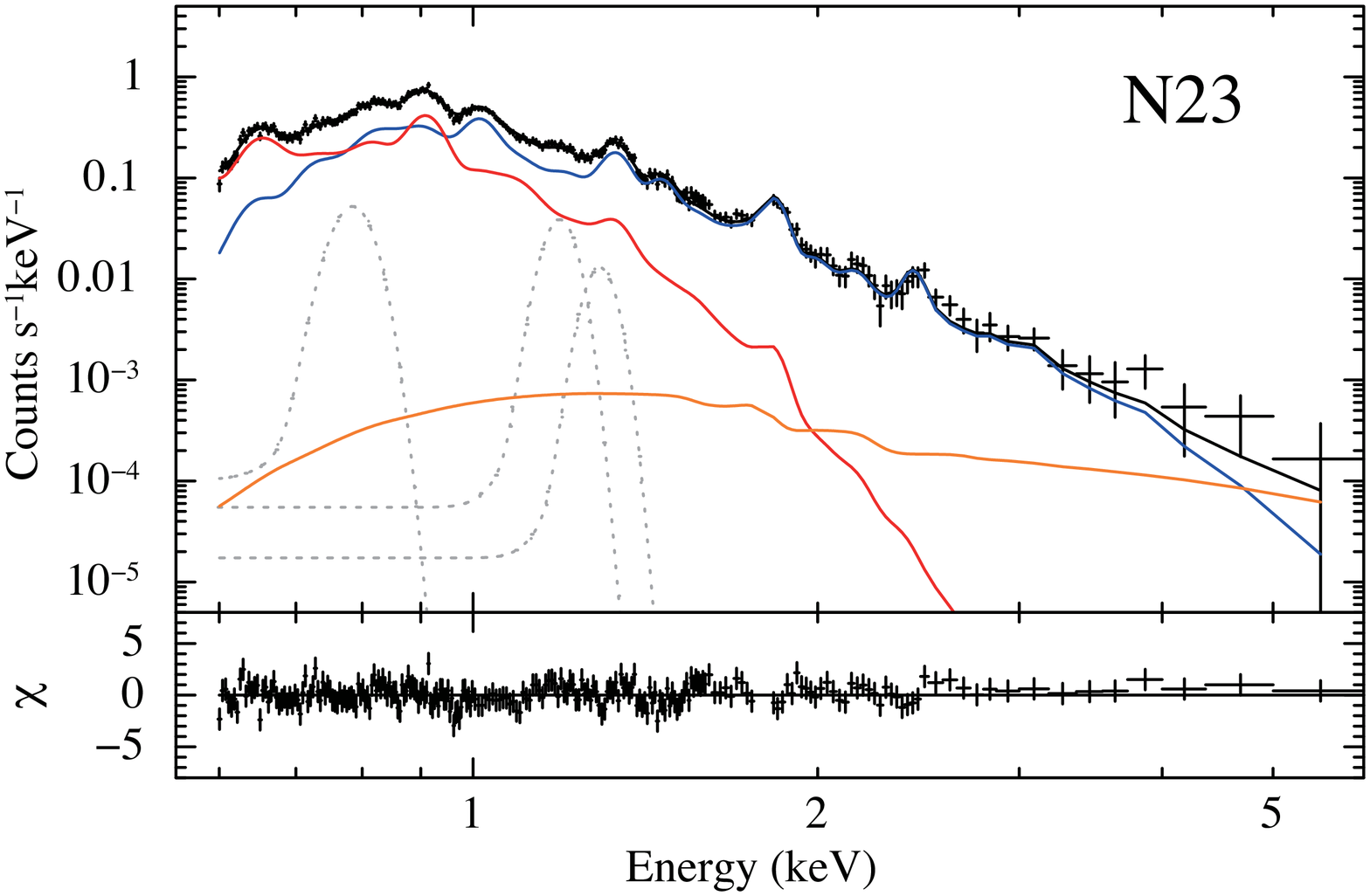}
 \includegraphics[width=70mm]{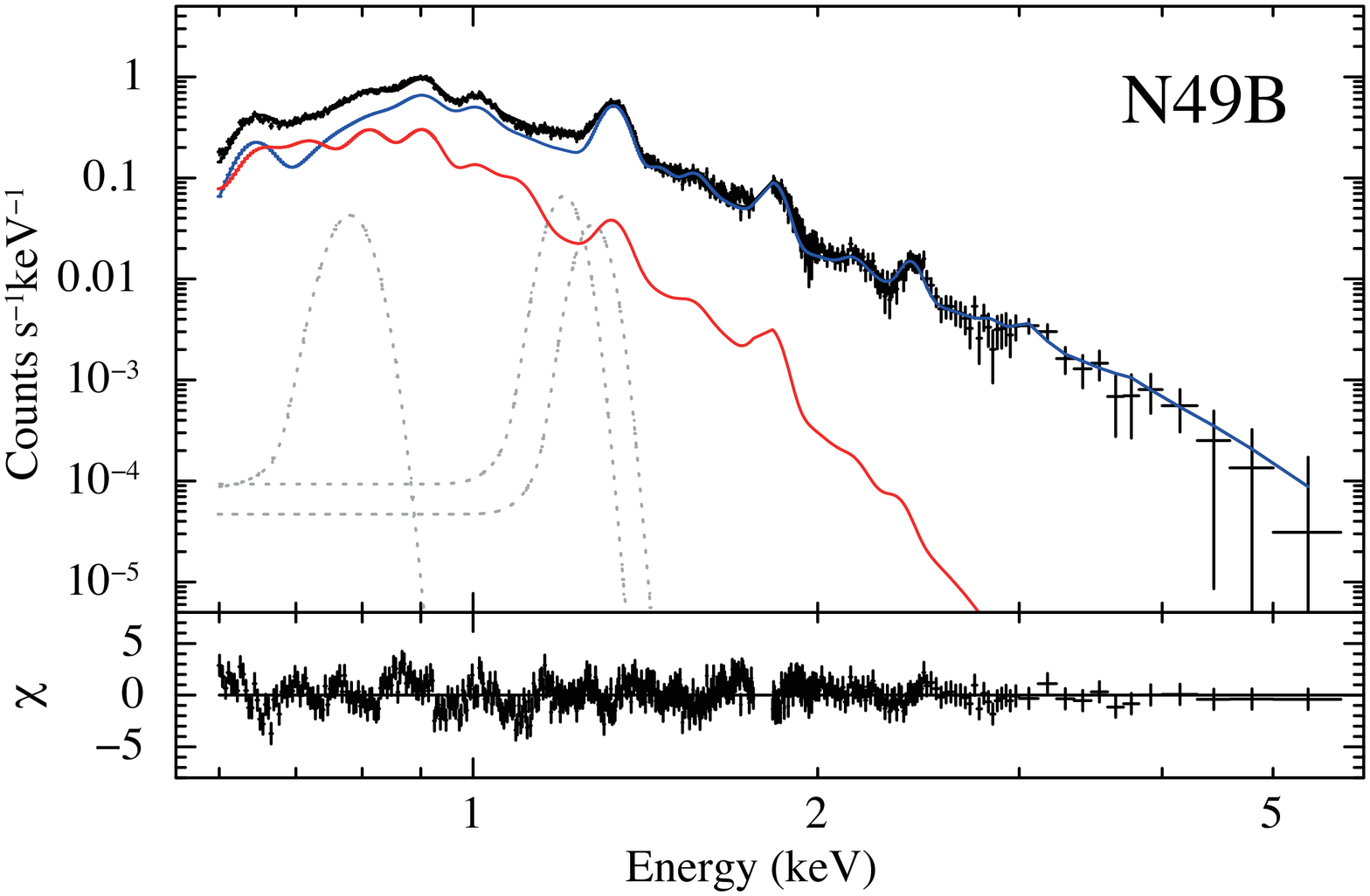}
 \includegraphics[width=70mm]{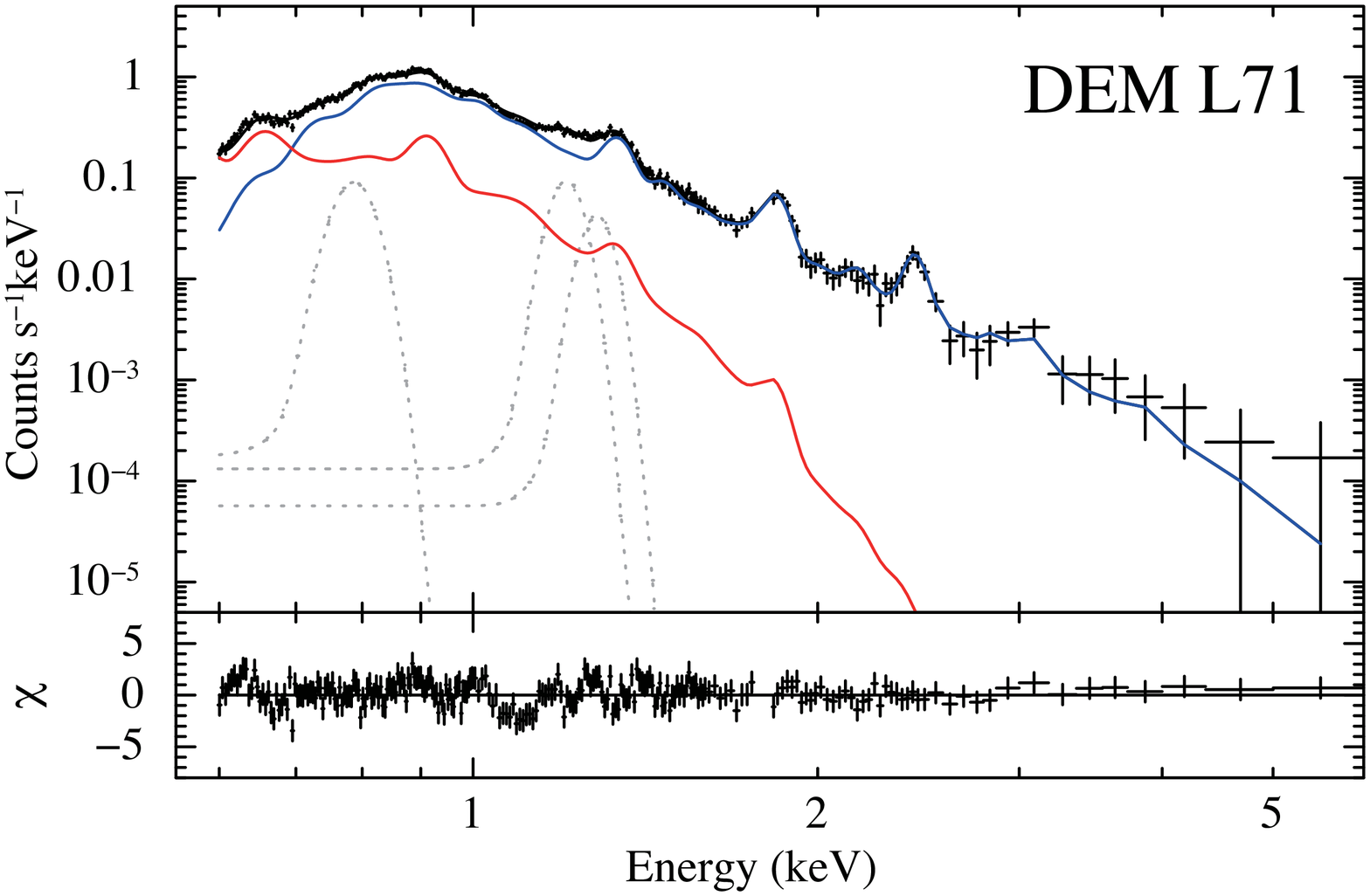}
 \end{center}
 \caption{XIS FI spectra of N23 (upper left), N49B (upper right) and DEM~L71 (bottom).
The best-fit models are overlaid with the black solid lines. The solid blue and red lines represent the best-fit high-$kT_e$
and the low-$kT_e$ components, respectively. The dotted lines show Gaussians for
the Fe L emission missing from the plasma code we used. The orange line in the model of N23 shows the power-law component for
CXOU~J050552.3$-$680141 (see text). }
 \label{fig:threeSNRs}
\end{figure*}

\subsection{N49}

N49 is one of the brightest SNRs in the LMC at the various wavelengths \citep[e.g.,][]{Long1981, Dickel1998}.
The radio continuum and H$\alpha$ emission exhibit a clear shell structure \citep{Vancura1992, Dickel1995}, 
whereas the X-ray image shows an irregular morphology brightest inside the radio shell in the southwest 
region \citep{Park2012}, which is the typical characteristic of a MM SNR. 
N49 is spatially overlapping with the soft gamma-ray repeater (SGR) 0526$-$66 \citep{Cline1982}, although the physical association between the SGR and SNR is under debate \citep{Gaensler2001, Badenes2009}.
 Using {\it Chandra} data, \citet{Park2012} revealed that the X-ray spectrum of SGR 0526$-$66 is best reproduced by a model consisting of a blackbody (BB) with $kT_{\rm BB} =
0.44\pm0.02$~keV and a power law (PL) with $\Gamma = 2.50^{+0.11}_{-0.12}$.
Unfortunately, our {\it Suzaku} data do not spatially resolve the SGR from the SNR.
We thus add the best-fit model by \citet{Park2012} to the model spectrum of N49, but allow its spectral parameters to vary within the reported statistical uncertainty. 
Since \citet{Park2012} found a significant long-term variability in the X-ray flux of the SGR, we also allow the fluxes of both BB and PL components to vary freely.

We first apply a single-component IP model where the initial ionization balance is dominated by the neutral state for all heavy elements. 
The free parameters are the electron temperature ($kT_e$), ionization timescale ($n_et$, where $n_e$ and $t$ are the electron number density and the elapsed
time since the gas was shock heated), emission measure ($EM=\int n_{\rm e}n_{\rm H} V$), and abundances of O, Ne, Mg, Si, S, Ar, Ca and Fe relative to the solar values of
\citet{Anders1989}.
The Ni abundance is linked to the value of Fe, while the abundances of the other elements are fixed to the LMC averages of \citet{Russell1992}.
This model yields the best-fit electron temperature of $\sim0.6$\,keV, but fails to reproduce the overall spectrum with an unacceptable $\chi^2$/d.o.f.\ of 4936/820. 
The residual is shown in panel~(a) of Figure~\ref{fig:N49}. 
The particularly large disagreement between the data and model is seen below $\sim$1\,keV, implying presence of another plasma component with a lower temperature.
Assuming that this component originates from a swept-up interstellar medium (ISM) that is dominant at the outermost region of the remnant \citep{Park2003N49, Park2012}, we added
another IP component with the abundances fixed to the LMC mean values.
Although this two-component model gives a slightly better fit ($\chi^2$/d.o.f.\ = 4297/817), some line-like residuals still remain as shown in panel~(b) of Figure~\ref{fig:N49}.
The residual at $\sim$0.8~keV, also observed in our previous studies of other objects \citep[e.g.,][]{Uchida2013, Nakashima2013}, is likely due to the well-known uncertainty in the emissivity ratio between the \ion{Fe}{18} L-shell emission of $3s \rightarrow 2p$ and $3d \rightarrow 2p$ \citep{Gu2007}.
On the other hand, the residuals at $\sim1.2$~keV and $\sim1.3$~keV are caused by L-shell emission from high quantum numbers ($n > 5$) missing from the plasma code SPEX \citep{Brickhouse2000}.
We compensate for these lines by adding three Gaussians at these energies, and obtain the improved fit ($\chi^2$/d.o.f.\ = 2366/815) with the best-fit $kT_e$ of $\sim0.6$~keV and $n_et$ of $\sim10^{13}$~cm$^{-3}$~s for the high-$kT_e$ component. 
The obtained ionization timescale indicates that this IP model actually represents a CIE plasma at the given temperature.

The fit left the largest residuals at 2.0~keV and 2.6~keV (see panel~(c) of Figure~\ref{fig:N49}), 
which correspond to the centroid energies of the Si Ly$\alpha$ and S Ly$\alpha$ emissions, respectively.
This suggests that the average charge of these elements is higher than that expected in a 0.6-keV CIE plasma. 
Furthermore, a hump-like feature is found in the residual around 2.7~keV, indicating enhancement of the RRC of Si, as was observed in the other RP SNRs \citep[e.g., IC\,443;][]{Yamaguchi2009}.
Therefore, we introduce a RP model (using an ``NEIJ'' model in SPEX\footnote{See http://www.sron.nl/files/HEA/SPEX/manuals/manual.pdf for details.}) for the high-$kT_e$ component.
In addition to the parameters given in the IP model, the RP model has another free parameter $kT_{\rm init}$ of higher than $kT_e$, which describes a recombining history by $n_et$ in the $kT_e$ plasma starting from the initial ionization temperature of $kT_{\rm init}$.
This model dramatically improves the fit ($\chi^2$/d.o.f. = 1270/813), removing the large residuals above 2~keV as shown in the panel~(d) of Figure~\ref{fig:N49}. The best-fit model
components and parameters are given in Figure~\ref{fig:N49} and Table~\ref{tab:para}, respectively.

Interestingly, the RP model reproduces the Fe K$\alpha$ emission at $\sim6.6$~keV as well, 
although free electrons in the 0.6-keV plasma are not energetic enough to excite K-shell electrons of Fe. 
This indicates that the observed Fe K$\alpha$ emission predominantly originates from cascade processes after the radiative recombination into the excited levels of He-like ions (i.e., Fe$^{25+}$ + $e^-$ $\rightarrow$ Fe$^{24+*}$).
We emphasize that this is why $kT_{\rm init}$ is constrained to the very high value ($\gtrsim10$~keV) so that a significant fraction of the Fe ions remains at the H-like state in the current plasma.
It should also be noted that the observed centroid energy $6629_{-26}^{+31}$~eV is consistent with the value expected from the best-fit RP model (6658~eV) but significantly lower than that for a typical CIE plasma at $kT_e = 5$--10\,keV ($\sim6680$~eV).
This is another piece of evidence for the RP; the forbidden and intercombination lines are likely to be enhanced by the recombination processes.

\begin{table*}[t]
  \begin{center}
 \caption{Best fit parameters}\label{tab:para}
     \begin{tabular}{llllll}
      \hline\hline
        Component & Parameter \\
      \tableline
       &  & N49  & N23 & N49B & DEM~L71 \\
      Absorption \ \ \ \ & $N_{{\rm H(LMC)}}$ ($\times10^{21}$~cm$^{-2}$)  & $3.51^{+0.01}_{-0.04}$ & $2.13\pm0.05$ & $2.30^{+0.01}_{-0.06}$ & $3.70^{+0.30}_{-0.50}$ \\
       & $N_{{\rm H(MW)}}$ ($\times10^{21}$~cm$^{-2}$) &  0.6 (fixed) & 0.6 (fixed) & 0.6 (fixed)  & 0.6 (fixed)\\
     ISM  & $kT_e$ (keV) &  $0.30\pm0.01$ &  $0.17\pm0.01$ &  $0.21\pm0.01$ &  $0.16\pm0.01$\\
                               &  $n_{\rm e}t$ ($\times10^{11}$~cm$^{-3}$\,s) & $3.49^{+0.09}_{-0.26}$ & $>10$ & $1.23^{+0.09}_{-0.07}$ & $>10$\\
                              &  $EM$ ($\times10^{58}$~cm$^{-3}$)  & $3.54^{+0.03}_{-0.01}$ & $17.6\pm0.3$ & $6.0\pm0.9$ & $19^{+5}_{-4}$\\
     Ejecta   &  $kT_e$ (keV)     &   $0.62\pm0.01$ & $0.61\pm0.01$ & $0.79\pm0.01$ & $0.69\pm0.01$\\
       			   &  $kT_{\rm init}$ (keV)   &  $11\pm1$ & 0.01 (fixed) & 0.01 (fixed) & 0.01 (fixed)\\
                            &  $n_{\rm e}t$ ($\times10^{11}$~cm$^{-3}$\,s)  &  $7.00^{+0.44}_{-0.02}$  &  $5.08^{+0.42}_{-0.37}$ &  $0.84\pm0.01$ &  $1.77\pm0.07$\\
                            & O & $1.52^{+0.04}_{-0.03}$ & $3.5\pm0.2$ & $0.63^{+0.03}_{-0.02}$ & $0.30^{+0.05}_{-0.04}$\\
                            & Ne & $0.98^{+0.03}_{-0.02}$ & $1.25\pm0.07$ & $0.45\pm0.01$ & $0.25\pm0.02$\\
                            & Mg & $0.82\pm0.02$ & $1.16\pm0.07$ & $1.24\pm0.02$ & $0.62\pm0.03$\\
                            & Si   & $1.32^{+0.05}_{-0.02}$ & $1.2\pm0.4$ & $0.34\pm0.02$ & $0.48\pm0.04$\\
                            & S    & $1.46^{+0.03}_{-0.11}$ & $1.25\pm0.08$ & $0.50\pm0.08$ & $1.0\pm0.2$\\
                            & Ar   & $1.3^{+0.1}_{-0.5}$ & $<3.1$ & $0.7\pm0.6$ & $1.5^{+3.1}_{-0.3}$\\
                            & Ca  & $1.5^{+0.2}_{-0.9}$ & (=Ar) & (=Ar) & (=Ar)\\
                            & Fe (=Ni) & $0.24\pm0.01$ & $0.46\pm0.02$ & $0.35\pm0.01$ & $0.62\pm0.01$\\
                            &  $EM$ ($\times10^{58}$~cm$^{-3}$)  & $1.73\pm0.01$ & $0.62\pm0.01$ & $0.79\pm0.01$ & $1.36\pm0.01$\\
                            \tableline
 & $\chi ^2$/d.o.f.   & $1270/813$ & $218/234$ & $659/423$ & $291/229$\\
      \tableline
    \end{tabular}
 \end{center}
\end{table*}

\subsection{N23}

In the previous work using {\it XMM-Newton} MOS and RGS data, \citet{Broersen2011} claimed 
that the X-ray spectrum of N23 can be best reproduced by a two-component NEI model of which 
one of the components was a RP with $kT_{\rm init} = 3.0$\,keV and $kT_e = 0.18$\,keV. 
We first fit the XIS spectrum fixing their best-fit parameters, but fail to reproduce the XIS spectrum ($\chi^2$/d.o.f. = 455/236), 
leaving large residuals around the Ne-Ly$\alpha$ emission and the \ion{Fe}{17} recombination edge at $\sim1.25$~keV. 
The fit is not improved even if we allow the abundances to vary. Therefore, we thaw the temperatures 
and ionization parameters (but restricting $kT_{\rm init}$ to be higher than $kT_e$), obtaining a significantly 
better fit with $\chi^2$/d.o.f. = 224/234 and $n_et > 10^{12}$\,cm$^{-3}$\,s for both components. 
This high $n_et$ value indicates that the plasmas are in nearly CIE independently from 
the initial ionization population (otherwise ionizing). 
In fact, we are not able to constrain $kT_{\rm init}$ with this model. 
Finally, we apply an IP model consisting of a LMC-abundance component (for ISM) and 
a free-abundance component (for ejecta), and obtain a slightly better fit ($\chi^2$/d.o.f. = 218/234) 
than the RP (or CIE) model. The best-fit parameters and models are given in Table~\ref{tab:para} and 
Figure~\ref{fig:threeSNRs}, respectively. 
During the analysis, we fix the spectral shape and flux of the faint X-ray compact source 
CXOU~J050552.3$-$680141 associated with N23 to the reported values of \citet{Hayato2006}, 
although its contribution is negligible at $\lesssim$\,3\,keV (see Figure~\ref{fig:threeSNRs}). 

\subsection{N49B}

N49B is a typical shell-like SNR containing no X-ray point-like source within the SNR shell \citep{Park2003N49B}.
We fit the XIS spectrum with the same two-component IP model applied to N49, obtaining the best-fit 
results given in Table~\ref{tab:para}. 
The $n_et$ value for the ejecta component ($\sim8\times10^{10}$~cm$^{-3}$\,s) is 
the lowest among the SNRs studied in this work. 
We reveal that the abundance of Mg is significantly higher than those of the other elements 
(e.g., $\rm{Mg/O}\sim2.0$, $\rm{Mg/Ne}\sim2.8$, $\rm{Mg/Si}\sim3.6$). 
\citet{Park2003N49B} previously found ``Mg-rich'' ejecta near the center of the remnant. 
Our result shows that the Mg enrichment is confirmed even in the integrated spectrum from the entire SNR. 
We find no evidence for a RP; if $kT_{\rm init} > kT_e$ is assumed, the model clearly fail to reproduce 
the spectrum with  $\chi^2$/d.o.f. = 1582/426. Therefore, we conclude that this SNR is dominated by IPs.

\subsection{DEM~L71}

DEM~L71 is a middle-aged shell-like SNR, where a double-shock morphology was observed by 
\textit{Chandra} observations \citep{Hughes2003}, suggesting the presence of both reverse-shocked 
ejecta and swept-up ISM. \citet{Hughes2003} also revealed that the X-ray spectra from the interior 
regions are dominated by strong Fe L emission. 
Similarly to the spectral analysis of N49B, we fit the XIS spectrum of DEM~L71 with 
a two-component IP model consisting of a low-$kT_e$ ISM and a high-$kT_e$ ejecta. 
The fit is acceptable with $\chi^2$/d.o.f. of 291/229 as is given in Table~\ref{tab:para}.
The Fe abundance is highest among the four SNRs, consistent with the previous studies 
\citep[e.g.,][]{Hughes1998, Hughes2003} where the Type~Ia origin of this SNR was suggested.
If we assume $kT_{\rm init} > kT_e$, an unacceptable fit ($\chi^2$/d.o.f. = 399/228) is obtained, 
ruling out a RP scenario for this remnant.

\section{Discussion}

\subsection{Plasma Conditions in the Observed SNRs}

\begin{table*}[t]
\caption{Summary of the Four LMC SNRs Studied in This Work}\label{tab:four-snrs}
\begin{center}
\begin{tabular}{lccccccc}
\hline
\hline
 SNR  & RP &Age ($10^4$~yr) & SN Type  & Morphology  & Cloud Interaction &References \\
\hline
\ \ N49  & Yes & 0.4--0.5 &  Core-collapse & MM & Yes (CO and H$\alpha$) &  this work, 1, 2, 3 \\
\ \ N23 &  No & 0.4 & Core-collapse & MM &  No  &   2, 4 \\
\ \ N49B & No & 1.09& Core-collapse & Shell  & No   &  5 \\
\ \ DEM~L71 & No & 0.4--0.5 & Ia & Shell   & No  &   6, 7, 8\\
\hline
\hline
\end{tabular}
\end{center}
\tablerefs{
(1) \citealt{Park2012};
(2) \citealt{Banas1997}; 
(3) \citealt{Melnik2013}; 
(4) \citealt{Broersen2011};
(5) \citealt{Park2003N49B}; 
(6) \citealt{Ghavamian2003}; 
(7) \citealt{Hughes1998}; 
(8) \citealt{Hughes2003}.}
\end{table*}

We have systematically analyzed X-ray spectra from the four LMC SNRs (N49, N23, N49B, and DEM~L71), 
and revealed the robust evidence for overionization in N49 --- this is the first discovery of a RP from 
an extra-galactic SNR. The other SNRs, including N23 from which presence of a RP was previously 
reported \citep{Broersen2011}, can be fairly well characterized by an IP or a nearly-CIE plasma.

The previous claim of the recombining state in N23 was based on the enhanced $G$-ratio of the \ion{O}{7} 
lines measured using the \textit{XMM-Newton} RGS. 
Although \citet{Broersen2011} claimed that the broadband spectrum of the \textit{XMM-Newton} MOS 
can also be reproduced by a RP model, their best-fit model is clearly ruled out by our analysis of 
the XIS spectrum with better photon statistics. 
Since our analysis excluded the \ion{O}{7} emission, it is still possible that 
the swept-up ISM (low-$kT_e$ component) is partially recombining. 
We note, however, that there are several other processes that 
can enhance the $G$-ratio: resonance scattering and charge exchange. 
Notably, a similar $G$-ratio enhancement in the \ion{O}{7} lines was observed in DEM~L71,  
and was indeed interpreted to be a consequence of the resonance scattering \citep{derHeyden2003}. 
Since the charge exchange process seems to work in some evolved SNRs \citep[e.g., Puppis~A;][]{Katsuda2012}, 
this process may also be responsible for the high $G$-ratio observed in N23.

\subsection{Origin of the Recombining Plasma}

We summarize the characteristics of the four SNRs studied in this work in Table~\ref{tab:four-snrs}. 
Only N49 is interacting with dense clouds identified by the CO \citep{Banas1997} 
and H$\alpha$ \citep{Melnik2013} observations.
Given that similar cloud interaction is observed in most SNRs from which the presence of a RP 
has been confirmed \citep[e.g., W49B;][]{Ozawa2009}, thermal conduction into surrounding clouds might play an important 
role in forming the RP \citep[see][for a discussion]{Zhou2011}. 
In this scenario, 
there should be a spatial correlation between the electron temperature 
of the SNR plasma and the cloud density. Future deep observations of N49 with better spatial resolution 
are necessary to assess this possibility.

An alternative is that the adiabatic expansion of the SNR had caused a rapid cooling of electrons and 
the resulting recombining state of the plasma. 
In this scenario, the SN ejecta 
should interact with a dense circumstellar matter (CSM) in the early phase of the SNR evolution, 
and the highly ionized ejecta and CSM expand drastically after the SNR blast wave breaks out to 
the low-density ISM  \citep[][]{Itoh1989}. 
Therefore, the ionization timescale ($n_et$) in this scenario is characterized by the elapsed time 
and the density evolution history since the break out took place. 
To evaluate the possibility of this scenario, we estimate the electron density in the present RP 
of N49 using the derived $EM$ (= $n_e n_{\rm H} V$). Since the plasma density is known to be highly 
inhomogeneous in this remnant \citep{Park2012}, we analyze archival {\it Chandra} data of N49 to 
investigate the surface brightness profile. We find that $\sim50$\% of the photon flux in the Si band (1.7--2.1\,keV) 
is coming from the brightest southeast region (approximately a $0.\!'4 \times 0.\!'25 \times 0.\!'25$ ellipsoid), 
and the other $\sim50$\% is from the remaining faint region in the entire SNR (a sphere with a radius of $0.\!'6$). 
Therefore, the volumes and the electron densities are estimated to be 
$9.6\times 10^{57}$\,cm$^3$ and 1.0\,cm$^{-3}$ for the bright region, and 
$7.3\times 10^{58}$\,cm$^3$ and 0.38\,cm$^{-3}$ for the faint region, at a distance of 50\,kpc. 
If we simply divide the best-fit $n_et$ in the ejecta component ($7.0 \times10^{11}$~cm$^{-3}$\,s)
by the derived electron density, we obtain plasma ages of 22--58\,kyr, 
more than a few times larger than the dynamical age of N49 \citep[$\sim$4800~yr;][]{Park2012}. 
This is not surprising because the plasma must have had a higher electron density in the past. 
\citet{Yamaguchi2012AdSpR} also obtained a similar result from 
the \textit{Suzaku} observation of IC~443, where the plasma age estimate ($\sim11$~kyr) is 
significantly higher than the SNR age of $\sim4000$~yr \citep{Troja2008}.

In Table~\ref{tab:rpsnr}, we summarize physical properties of the RP SNRs identified so far.
While the estimated ages are roughly correlated with $kT_e$, the values of $n_et$ 
show no clear trend. This may be partly due to a technical reason that the parameter $n_et$ 
is coupled to that of $kT_{\rm init}$ in the spectral fitting.
Nevertheless, Table~\ref{tab:rpsnr} shows that the $n_et$ values exceed $\sim10^{11}$~cm$^{-3}$\,s  
in all SNRs, implying that the overionization did \textit{not} start from the last few hundred years, but from the early epoch in the SNR.
This fact likely favors the adiabatic cooling scenario for the most RP SNRs.

\begin{table*}[t]
\caption{List of the RP SNRs in the Order of Increasing Electron Temperature}\label{tab:rpsnr}
\begin{center}
\begin{tabular}{lccccccc}
\hline
\hline
 SNR  & Age & $kT_{\rm init}$  & $kT_e$  & $n_et$ & Cloud Interaction & Compact Object & References \\
&    ($10^4$~yr) & (keV) & (keV) & ($10^{11}$~cm$^{-3}$\,s) & & \\
\hline
\ \ G359.1$-$0.5 &  $>1$ & $0.87^{+0.15}_{-0.11}$ & $0.29\pm0.02$ &  $<4.42$  & Yes & \nodata & 1, 2, 24 \\
\ \ G346.6$-$0.2  & \nodata& 5 (fixed) & $0.30^{+0.03}_{-0.01}$ & $4.8^{+0.1}_{-0.4}$  & Yes &  \nodata & 3, 25 \\
\ \ W28 &  $\sim4$ & 3 (fixed) & $0.40^{+0.02}_{-0.03}$ & 6.31  & Yes & PSR~1758$-$23 & 4, 5, 6, 24\\
\ \ W44 &  $\sim2$ & $1.07^{+0.08}_{-0.06}$ & $0.48\pm0.02$ & $6.76\pm0.5$ & Yes &  PSR~B1853$+$01  & 7, 8, 9, 24 \\
\ \ CTB~37A &  $\sim1$ & 5 (fixed) & $0.49^{+0.09}_{-0.06}$ & $13^{+3}_{-1}$  & Yes & CXOU~J171419.8$-$383023  & 10, 11, 12, 24 \\
\ \ 3C391   &\nodata& 1.8$^{+1.6}_{-0.6}$ & 0.495$\pm0.015$ & $14.0^{+1.5}_{-2.2}$  & Yes & \nodata  & 13, 24  \\
\ \ MSH~11-61A  & 1--2 & 5 (fixed) & $0.513^{+0.004}_{-0.003}$ & $12.2\pm0.4$ & Possible & IGR J11014-6103  & 14, 15, 16, 26\\
\ \ N49  & 0.4--0.5 &  $11\pm1$ & $0.62\pm0.01$ & $7.00^{+0.44}_{-0.02}$ & Yes &  SGR~0526$-$66 & this work, 17 \\
\ \ IC~443   & $\sim0.4$ & 10 (fixed) & $0.65\pm0.04$  & $9.8\pm1.1$  & Yes &   XMMU~J061804.3$+$222732  & 18, 19, 20, 21, 24\\
\ \ W49B   & $\sim0.4$ & \nodata & $1.52^{+0.01}_{-0.02}$  & \nodata & Yes  & \nodata & 22, 23, 27 \\
\hline
\hline
\end{tabular}
\tablecomments{The plasma parameters for IC~443 and W49B were determined only from a high-energy band around Fe-K lines \citep[e.g.,][]{Ohnishi2014, Ozawa2009}, while those for the other SNRs were determined from intermediate-mass elements.}
\end{center}
\tablerefs{
(1) \citealt{Aharonian2008HESSJ1745-303};
(2) \citealt{Ohnishi2011}; 
(3) \citealt{Yamauchi2013}; 
(4) \citealt{Rho2002W28};
(5) \citealt{Sawada2012};
(6) ``runaway'' Pulsar; \citealt{Frail1993}; 
(7) \citealt{Cox1999}; 
(8) \citealt{Uchida2012W44}; 
(9)  Pulsar plus Pulsar Wind Nebula; \citealt{Petre2002};
(10) \citealt{Yamauchi2014};
(11) \citealt{Wolszczan1991};
(12)  Pulsar Wind Nebula candidate; \citealt{Aharonian2008CTB37A}; 
(13)  \citealt{Sato2014}; 
(14) \citealt{Slane2002}; 
(15) \citealt{Kamitsukasa2015}; 
(16) Pulsar plus Pulsar Wind Nebula; \citealt{Pavan2014};
(17) \citealt{Park2012}; 
(18) \citealt{Lee2008}; 
(19) \citealt{Yamaguchi2009}; 
(20) \citealt{Ohnishi2014}; 
(21) Neutronstar plus Synchrotron Nebula; \citealt{Olbert2001};
(22) \citealt{Hwang2000};
(23) \citealt{Ozawa2009};
(24) \citealt{Hewitt2008};
(25) \citealt{Koralesky1998};
(26) \citealt{Filipovic2005};
(27) \citealt{Keohane2007}.}
\end{table*}

\subsection{Progenitor of N49 and SGR\,0526--66}

The SN type of N49 has been controversial. \citet{Park2012} suggested a Type Ia origin
based on the Si/S abundance ratio in the ejecta measured by the {\it Chandra} data. 
On the other hand, the environment of N49, i.e., a nearby OB association \citep{Chu1988} 
and young stellar clusters \citep{Klose2004}, are more common in core-collapse SNRs. 
A recent systematic study of Fe K emissions in young and middle-aged SNRs also agrees 
on its core-collapse origin \citep{Yamaguchi2014}.
Given that a RP has never been observed in a Type Ia SNR, our results strongly support 
a core-collapse scenario for this SNR. It should also be noted that 
the formation of a RP needs dense CSM, which is only made by 
a massive progenitor with a significant mass 
loss rate \citep{Moriya2012, Shimizu2012}. 
The total mass of the RP component is roughly estimated to be $\sim$26$M_{\odot}$ 
using the density and emitting volume derived in \S4.2, 
which is in marginal agreement with the progenitor mass of 12.5--21.5\,$M_{\odot}$ 
constrained from the local star formation history around N49 \citep{Badenes2009}. 

Since the core-collapse origin is strongly suggested for N49, the compact object SGR~0526$-$66, 
which is known to be a magnetar candidate \citep[e.g.,][]{Park2012}, might be associated with the SNR. 
The large distance between SGR~0526$-$66 and the geometric center of N49 requires 
a high kick velocity ($\sim1100$~km~s$^{-1}$) for the SGR at the SNR age of $\sim4800$~yr, and hence \citet{Park2012} argued against the physical association between these objects. 
However, the theoretical prediction is that a magnetorotational core-collapse SN explodes asymmetrically with 
bipolar jets and produces a magnetar with a large kick velocity up to $\sim1000$~km~s$^{-1}$ \citep{Sawai2008}.
Moreover, there are several compact objects located extremely off-center from the associated RP SNRs.
For instance, IGR~J11014$-$6103 is located outside the shell of MSH~11-61A.
\citet{Pavan2014} indicated that this neutron star is escaping from MSH~11-61A at a velocity exceeding 
1000~km~s$^{-1}$. Also, PSR~1758$-$23 is known as a ``runaway pulsar'' which might be physically 
associated with W28 \citep{Frail1993}. Two other compact objects, CXOU~J171419.8$-$383023 in 
CTB~37A \citep{Aharonian2008CTB37A} and XMMU~J061804.3$+$222732 in IC~443 \citep{Olbert2001}, 
are also found far from the geometric center of their host SNRs.
Taking these facts into consideration, we propose that SGR~0526$-$66 is the plausible candidate of 
the compact remnant of N49.

\section{Conclusions}

We have shown the clear evidence for the RP in the SNR N49, which is the first report of the robust detection of a RP from an extra-galactic SNR achieved by the high sensitivity and the good energy resolution of the XIS on board {\it Suzaku}. The other SNRs, including N23 where presence of a RP was previously claimed (Broersen et al.\ 2011), can be well characterized by an IP or a nearly-CIE plasma. A future observation of N49 using high-resolution X-ray spectrometers, like {\it ASTRO-H} \citep{Takahashi2014}, is highly encouraged; our {\it Suzaku} result predicts the detection of the Fe RRC as well as the enhanced forbidden lines due to the recombination processes (see \S3.1). We also emphasize the importance of a systematic study of the RP SNRs, which will help us understand physical processes responsible for the formation of the RPs in more details.

\acknowledgments
The authors thank to Dr. T. G. Tsuru for a careful reading of our manuscript.
This work is supported by JSPS KAKENHI Grand Numbers 26800102 (H.U.) and 2450229 (K.K.).

{\it Facilities:} \facility{\textit{Suzaku} (XIS)}.

\bibliographystyle{apj}
\bibliography{References}

\end{document}